# Astro2020 Science White Paper

# Identification and characterization of the host stars in planetary microlensing with ELTs

**Thematic Areas:**  ☒ Planetary Systems  ☐ Star and Planet Formation

☐ Formation and Evolution of Compact Objects  ☐ Cosmology and Fundamental Physics

☐ Stars and Stellar Evolution  ☐ Resolved Stellar Populations and their Environments

☐ Galaxy Evolution  ☐ Multi-Messenger Astronomy and Astrophysics

**Principal Author:**
Name: Lee, Chien-Hsiu
Institution: National Optical Astronomy Observatory
Email: lee@noao.edu
Phone: 5203188000

**Co-authors:** (names and institutions)
Rachel Street, LCOGT
Kailash Sahu, STScI
Eliad Peretz, NASA/GSFC

**Abstract** (optional):
Microlensing offers a unique opportunity to probe exoplanets that are temperate and beyond the snow line, as small as Jovian satellites, at extragalactic distance, and even free floating exoplanets, regimes where the sensitivity of other methods drops dramatically. This is because microlensing does not depend on the brightness of the planetary host star. The microlensing method thus provides great leverage in studying the exoplanets beyond the snow line, posing tests to the core accretion mechanism, especially on the run-away phase of gas accretion to form giant planets. Here we propose to robustly and routinely measure the masses of exoplanets beyond 1 AU from their host stars with the microlensing method; our experiment relies on directly imaging and resolving the host star (namely the lens) from the background source of the microlensing events, which requires the high spatial resolution delivered by the ELTs. A direct result from this project will be planet occurrence rate beyond the snow line, which will enable us to discern different planet formation mechanisms.

Microlensing provides a unique way to probe exoplanets that are otherwise too faint or too far away from their host stars to be detected by transit, radial velocity, and/or direct imaging (see Fig. 1). This is because the microlensing method depends neither on the brightness of the exoplanets nor that of the host stars, but only on the gravitational field exerted by the exoplanets. In this regard, microlensing can probe sub-Neptune and even Earth mass exoplanets beyond the snow line (see e.g. Gaudi 2012), as well as free-floating planets that are not associated with any host stars (see e.g. Mroz et al. 2017). In addition, as most of the exoplanets from radial velocity, transit, and direct imaging method are within hundred of parsecs, microlensing is the only method that can detect exoplanets at several kpcs, making it the ideal technique with which to explore the full population of exoplanets in the Galaxy. It has the potential to discover exoplanets in the Galactic plane (see e.g. Sajadian & Poleski 2018), in particular the new surveys such as ZTF and LSST will provide microlensing event discoveries in substantial numbers across a wider area of the Galactic Plane, visible from both North and South hemispheres. Microlensing can even discover exoplanets at extragalactic distance, e.g. in the Small Magellenic Cloud (Di Stefano 2000) and in the Andromeda galaxy (Ingrosso et al. 2009). Microlensing provides tremendous leverages to study planet formation and evolution in environments beyond the reach of other exoplanet detection methods.

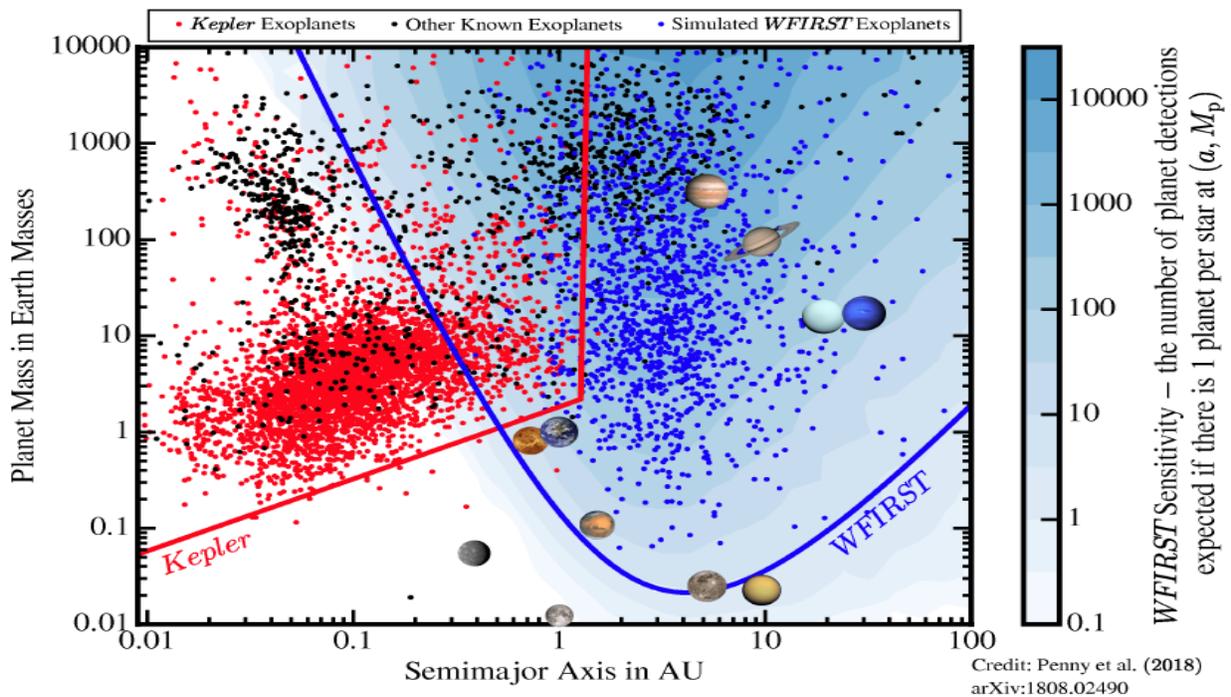

Figure 1: Predictions of microlensing exoplanets detection sensitivity by a dedicated WFIRST survey of the Galactic bulge, by Penny et al. (2018).

However, the microlensing light curve only provides the information about the planet-to-star mass ratio, the projected separation between the planet and the star, and event time scale, which depends on a combination of the lens mass, the lens-source relative proper motion, and the distance to the lens system. Thus, we will need further constraints to determine the absolute mass and orbits of the exoplanets (see e.g. Lee 2017). One possibility is to use high resolution imaging, which can resolve the lens from the source -- once the lens-source relative motion causes them to separate on the plane of the sky -- and provides the lens-source relative proper motion and photometry of the lens (Fig. 2). With the lens brightness in hand, we can use the mass-luminosity relation to determine the primary lens mass and in turn infer the companion planetary mass without ambiguity. The lens-source relative proper motion, when coupled with the distance of the lens, can provide additional constraints on the lens mass as well. There have been dedicated high spatial resolution HST or AO imaging from Keck/VLT/Subaru to extract the light contribution of planetary microlenses for a handful of events from current microlensing surveys, in particular the OGLE, MOA, and KMTNet (see e.g. Batista et al. 2015). However, the next generation microlensing surveys, e.g. PRIME, Euclid, and in particular WFIRST (targeting i ~19-25 mag stars) will deliver > 1000 microlensing planets that is beyond the reach of comprehensive imaging using 8-m telescopes. To have a comprehensive understanding of the planetary microlensing events, we will need exquisite imaging of a great portion of the Milky Way bulge routinely monitored by the microlensing surveys, which can only be delivered by ELTs in a timely and efficient manner. Beyond that, it is particularly important to perform AO imaging for short-timescale lensing events, which are most probably caused by the lowest mass lenses, including free-floating planets, since other methods of constraining the mass (e.g. measuring parallax) cannot be applied in these cases.

We should note that besides direct imaging, there is another possibility of using microlensing parallax to break the mass-distance degeneracies (see. e.g. Gould 2004, Gould et al. 2009, Street et al. 2016). However, this method requires simultaneous observations by large telescope (i.e. highly competitive) resources (e.g. WFIRST + LSST) of events while in progress -- which is both expensive and harder to schedule. In contrast, ELT observations can effectively be done at leisure, several years after the event.

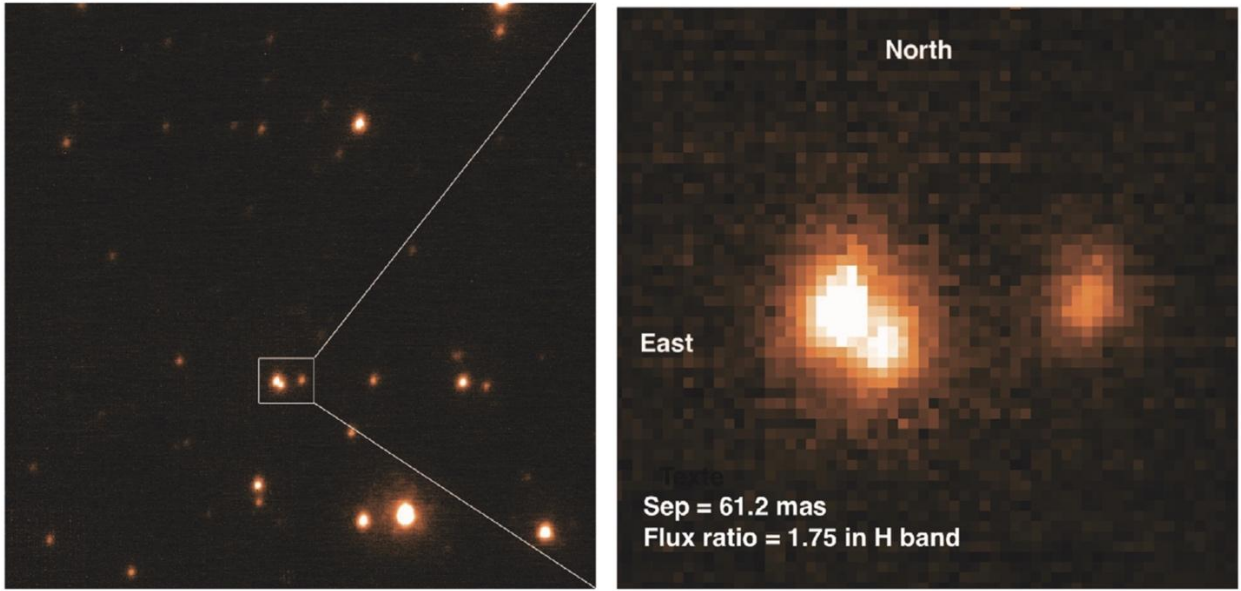

Fig. 2. Left panel: Kech H-band imaging of microlensing event OGLE-2005-BLG-169, taken at 8.21 years after the peak magnification of the microlensing light curve. Right panel: Zoom-in of the left panel, showing the lens on the upper left, separated by 61 milli-arcseconds from the source in the middle, with a faint 3rd light interloper on the right. Image from Batista et al. (2015).

The host stars of the planetary microlensing are rather faint, with i ~ 19-25 mag, and will require 5-10 minutes on-source AO imaging per filter with ELTs to identify the planetary host. This amounts to ~ 50 nights to survey all microlensing events (from WFIRST alone). While it can be done with 8-10m telescopes, we will need to wait longer time-span (~ 10 years after the microlensing event) to resolve the lens from the source, given the inferior spatial resolution of 8m class telescopes, and it demands much more exposure time on an 8-m telescope than ELTs to complete the survey. For estimate, the vast majority of the microlensing event has relative source-lens proper motion of 2-6 milli-arcsecond per year (e.g. Calchi Novati et al. 2018), hence 10 years must elapse before an 8m-class facility with spatial resolution of 0.01 arcsec/pixel with AO corrected seeing of 0.06 arcsec will be able to distinguish the stars. In contrast, ELT need wait only 5 years to separate the stars with the superior resolution of 0.004 arcsec/pixel and AO corrected seeing of 0.02 arcsec.

While WFIRST will conduct a dedicated Galactic bulge microlensing survey, there are more and more microlensing events being picked up by all-sky surveys, especially toward the Galactic plane and other dense stellar region. In this regard, LSST will also serve as another source of microlensing, particularly outside the Bulge (see e.g. Gould 2013). This requires northern hemisphere follow-up, since it will probe deep enough to detect events across the sky. For example, TCP J05074264+2447555 b (Nucita et al. 2018, Dong et al. 2018) is an exoplanetary microlensing towards the Taurus region in the northern hemisphere, hence both TMT and GMT will be essential to cover microlensing events from both hemisphere. This will enable us to explore the distribution of otherwise unseen lensing populations throughout the Galaxy (including stellar remnant lenses). We note that while Mroz et al. (2017) suggested that ultra-short microlensing events may indicate the existence of free-floating Earth-mass exoplanets, new analysis by Niikura et al. (2019) point out that these ultra-short microlensing can also be reproduced by Earth-mass promidal black holes. In this regard, not detecting any light from a lens with an ELT places a stronger constraint on possible single black hole lenses -- which no other technique can detect.

It is only with precise mass measurement of a large sample of microlensing events in hand, that we can conduct statistical studies of mass distribution and planet occurrence rate beyond the snow line using microlensing events, which provides a stringent constraint on the planet formation theory (Gould et al. 2010, Cassan et al. 2012). This situation is analogous to that of exoplanetary transits, where using high resolution spectroscopy Fulton et al. (2017) identified two distinct exoplanet populations of super Earths and sub Neptunes, which was only possible with the precise radius measurement from ~ 5,000 exoplanets. To investigate the mass distribution we will also need a sample with similar number of sources. Currently there are only ~ 200 mass measurements for exoplanets, and the microlensing method, with WFIRST surveys, can routinely determine the exoplanets mass for ~ 1,000 exoplanets.

We note that there has been investigation of planet occurrence rate beyond the snow line using microlensing event (see Fig. 3), however, with large uncertainty due to the small number statistics. With the microlensing mass measurements from ELTs, we will be able to put stringent constraints. In addition, there are several lines of evidences from microlensing suggesting that sub-Saturnian mass exoplanets are (~ 10x) more numerous than the core accretion model predicted (Suzuki et al. 2018, Bhattacharya et al. 2018). These results, while preliminary, demonstrate that the microlensing provides a promising channel to study the planet formation mechanism beyond the reach of other planet detection methods.

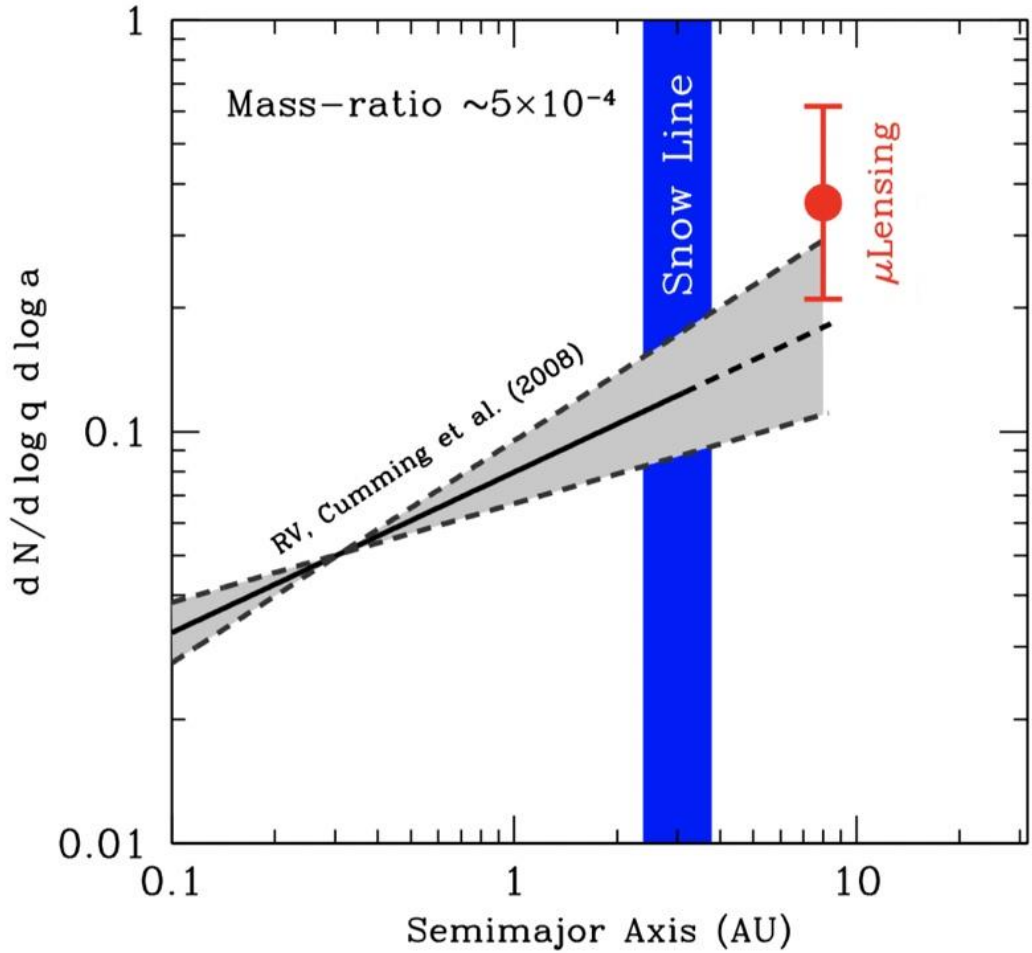

Fig. 3. Planet occurrence rate from radial velocity surveys (Cumming et al. 2008) and microlening (Gould et al. 2010). Microlensing is currently the only method to conduct comprehensive statistic studies of planets beyond the snowline. Image from Gould et al. (2010)